\documentclass[prl,aps,superscriptaddress,twocolumn]{revtex4-1}
\usepackage{amsmath,amssymb}
\usepackage{graphicx}
\usepackage{wasysym}
\usepackage{amsfonts}
\usepackage{bm}
\usepackage{enumerate}
\usepackage{color}
\usepackage[resetlabels]{multibib}
\usepackage{epstopdf}
\usepackage{latexsym}
\usepackage[breaklinks,colorlinks = true,linkcolor = red,urlcolor=cyan,citecolor=red]{hyperref}
\usepackage[caption=false,singlelinecheck=false]{subfig}

\usepackage{times}
\newcommand{\bea}{\begin{eqnarray}}
\newcommand{\eea}{\end{eqnarray}}
\newcommand{\be}{\begin{eqnarray}}
\newcommand{\ee}{\end{eqnarray}}
\newcommand{\bw}{\begin{widetext}}
\newcommand{\ew}{\end{widetext}}

\newcommand{\bs}{\boldsymbol}

\begin{document}
\title{Multipolar superconductivity in Luttinger semimetals}
\author{GiBaik Sim}
\email{gbsim1992@kaist.ac.kr}
\affiliation{Department of Physics, Korea Advanced Institute of Science and Technology, Daejeon 305-701, Korea}
\author{Archana Mishra}
\email{amishra@kaist.ac.kr}
\affiliation{Department of Physics, Korea Advanced Institute of Science and Technology, Daejeon 305-701, Korea}
\affiliation{International Research Centre MagTop, Institute of Physics, Polish Academy of Sciences,\\ Aleja Lotnik\'ow 32/46, PL-02668 Warsaw, Poland}
\author{Moon Jip Park}
\email{moonjippark@kaist.ac.kr}
\affiliation{Department of Physics, Korea Advanced Institute of Science and Technology, Daejeon 305-701, Korea}
\author{Yong Baek Kim}
\email{ybkim@physics.utoronto.ca}
\affiliation{Department of Physics, University of Toronto, Toronto, Ontario M5S 1A7, Canada}
\affiliation{School of Physics, Korea Institute for Advanced Study, Seoul 02455, Korea}
\author{Gil Young Cho}
\email{gilyoungcho@postech.ac.kr}
\affiliation{Department of Physics, POSTECH, Pohang, Gyeongbuk 790-784, Korea}
\author{SungBin Lee}
\email{sungbin@kaist.ac.kr}
\affiliation{Department of Physics, Korea Advanced Institute of Science and Technology, Daejeon 305-701, Korea}

\date{\today}

\begin{abstract}
Topological superconductivity in multiband systems has received much attention due to a variety of possible exotic superconducting order parameters as well as non-trivial bulk and surface states. While the impact of coexisting magnetic order on superconductivity has been studied for many years, such as ferromagnetic superconductors, the implication of coexisting multipolar order has not been explored much despite the possibility of multipolar hidden order in a number of $f$-electron materials. In this work, we investigate topological properties of multipolar superconductors that may arise when quadrupolar local moments are coupled to conduction electrons in the multiband Luttinger semimetal. We show that the multipolar ordering of local moments leads to various multipolar superconductors with distinct topological properties. We apply these results to the quadrupolar Kondo semimetal system, PrBi, by deriving the microscopic multipolar Kondo model and examining the possible superconducting order parameters. We also discuss how to experimentally probe the topological nature of the Bogoliubov quasiparticles in distinct multipolar superconductors via doping and external pressure, especially in the context of PrBi.
\end{abstract}

\maketitle

One of the foremost themes in contemporary condensed matter physics is the realization of topological superconductivity (TSC), where Bogoliubov-de Gennes (BdG) quasi-particles are characterized by non-trivial topology\cite{qi2009time,qi2010topological,hasan2010colloquium,sato2017topological}. Among the numerous proposals to realize the TSCs\cite{fu2008superconducting,chung2011conductance,cho2012superconductivity,bednik2015superconductivity,li2018topological,nadj2014observation,he2017chiral}, a promient route is to utilize multiband or multi-orbital superconductivity\cite{brydon2016pairing,agterberg2017bogoliubov,timm2017inflated,brydon2018bogoliubov,menke2019bogoliubov,roy2019topological,szabo2018interacting,boettcher2018unconventional,herbut2019ground,sim2019topological,venderbos2018pairing,savary2017superconductivity,yang2016topological,wu2010quintet,tchoumakov2019superconductivity,continentino2014topological,greg2019topo,kriener2011bulk,deng2012majorana,kawakami2018topological,sim2019triplet}, where the Cooper pairs possess non-zero angular momentum through the interband pairing channels. A representative example is the superconductivity in pseudospin $j=3/2$ Luttinger semimetals\cite{luttinger1955motion,luttinger1956quantum} with low-energy excitations described by quadratic band touching.

The multiband nature of the Luttinger semimetals has motivated intensive research on the possible unconventional superconductors
supporting the Coopr pairs with higher pseudospin angular momentum $j$ \cite{brydon2016pairing,agterberg2017bogoliubov,brydon2018bogoliubov,timm2017inflated,menke2019bogoliubov,roy2019topological,szabo2018interacting,boettcher2018unconventional,herbut2019ground,sim2019topological,venderbos2018pairing,savary2017superconductivity,yang2016topological}. 
In particular, it has been shown that the electron-electron interaction favors the $d$-wave pairing channels in the $j$ = 2 manifold 
over the $s$-wave in the $j$ = 0 state\cite{boettcher2018unconventional}.
Such unconventional superconductors possess a number of striking features including the emergent topological boundary states and the Bogoliubov Fermi surfaces with non-trivial Chern numbers.\cite{agterberg2017bogoliubov,timm2017inflated,brydon2018bogoliubov,oh2019instability}
All these interesting properties arise uniquely in multiband systems and result from the interplay between 
spin-orbit coupling and inter-band pairing channels.\cite{agterberg2017bogoliubov,venderbos2018pairing}
Among various candidate materials, a half-Heusler compound, YPtBi, shows the linear temperature dependence of London penetration depth\cite{kim2018beyond}, indicating the existence of unconventional nodal line superconductivity. In addition, other half-Heusler compounds such as LuPdBi and LaBiPt also exhibit superconductivity\cite{goll2008thermodynamic,nakajima2015topological}.
These half-Heusler compounds have negligible anisotropies of the Fermi surface near the quadratic band touching point\cite{meinert2016unconventional,oguchi2001electronic}. Therefore, the Luttinger model with $SO(3)$ or cubic symmetries 
have been employed to explain the superconductivity in these materials.\cite{brydon2016pairing,roy2019topological,boettcher2018unconventional,savary2017superconductivity}

On the other hand, unlike the half-Heuslers addressed above, other series of half-Heuslers like TbPdBi and HoPdBi exhibit unconventional superconductivity coexisting with magnetic ordering from rare-earth ions Tb and Ho.\cite{xiao2018superconductivity,radmanesh2018evidence} 
These materials are extremely interesting platforms for the study of the interplay between the magnetic degrees of freedom 
and unconventional superconductivity in multi-orbital systems. Furthermore, the pyrochlore oxide Cd$_2$Re$_2$O$_7$ and Pr based intermetallic compounds $\text{Pr}\text{(TM)}_\text{2} \text{X}_{20}$ (TM=Ti,V,Rh,Ir and X=Al,Zn) have recently been found to show coexistence of multipolar order and superconductivity.\cite{hanawa2001superconductivity,huang2009electronic,harter2017parity,matsubayashi2018high,ishii2013antiferroquadrupolar,onimaru2010superconductivity,onimaru2012simultaneous} 
For example, $\text{Pr}\text{(TM)}_\text{2} \text{X}_{20}$ systems show superconductivity 
near and below the temperature where the multipolar ordering is developed \cite{onimaru2010superconductivity,sakai2012superconductivity,onimaru2011antiferroquadrupolar,tsujimoto2014heavy,sato2012ferroquadrupolar,matsubayashi2012pressure}. 
Another semimetallic system, PrBi, is known to have both the quadrupolar degrees of freedom coming from Pr ions
and the $j$ = 3/2 Luttinger semimetal.
Recent experiments on this material have confirmed the existence of ferro-quadrupolar order originating from the localized 
moments of Pr ions, which may indicate the importance of the quadrupolar Kondo effect\cite{he2019prbi}.
Such situation is analogous to ferromagnetic superconductors, where the presence of magnetism can significantly alter
the nature of the superconducting state. Hence it is conceivable that the presence of multipolar order could
change the nature of the resulting multipolar superconductors in some fundamental ways. 

In this paper, motivated by the intertwined physics of multipolar order and superconductivity, we discuss how their coexistence can give rise to multipolar superconductivity with unique topological properties. In particular, we consider PrBi system as a concrete example and derive
the microscopic quadrupolar Kondo model, where the non-Kramers doublet of the localized Pr moments and 
the Bi itinerant electrons described by the Luttinger model are interacting with each other. 
In the absence of quadrupolar order, we first discuss the superconducting phases within the cubic symmetric Luttinger model. 
We find that time-reversal-symmetry breaking $d$-wave superconductors occur in the weak coupling limit, while the time-reversal symmetry is restored in the strong coupling limit. In the presence of quadrupolar order, however, we find that the superconducting instabilities are significantly altered in the way that the quadrupolar order induces Fermi surface distortion and stabilizes the multipolar superconductivity with mixtures of distinct $d$-wave pairing order parameters. Moreover, we find that these superconducting phases harbor topologically non-trivial gapless nodal line or nodal surface excitations, the nature of which sensitively depending on the quadrupolar order. 
Thus, one could change the topological properties of the multipolar superconductors by controlling the coexisting multipolar oder.
This would be a good example of magnetic topological phases that could be controlled by magnetism.
Based on our theory, we also propose various experiments that can probe the topological nature of the Bogoliubov quasiparticles 
in multipolar superconductors, anticipating potential applications to PrBi materials with doping and external pressure.

{\bf \em Luttinger model and electron interaction ---}
We start by describing the kinetics of the itinerant electrons with the Luttinger-semimetal Hamiltonian, 
\bea
H_0(\bm k)\!=\!c_0 k^2\!+\!\sum_{i=1}^{5}c_i d_i(\bm k)\gamma_i\!-\!\mu,
\label{eq:LH_c}
\eea
in four component spinor basis defined as $\psi\!\!\equiv\!\!(\psi_{3/2}, \psi_{1/2}, \psi_{-1/2}, \psi_{-3/2})$ and with the five $4\!\!\times\!\!4$ anti-commuting Dirac matrices, $\gamma_i$.\cite{boettcher2016superconducting}  Here, $\mu$ is the chemical potential, $d_i(\bm k)$ represent the five real $l\!\!\!=\!\!\!2$ spherical harmonics with $d_1\!\!\!=\!\!\!\sqrt{3}(k^2_x-k^2_y)/2$, $d_2(\bm k)\!\!=\!\!(3k^2_z-\bm k^2)/2$, $d_3(\bm k)\!\!=\!\!\sqrt{3}k_yk_z$, $d_4(\bm k)\!\!=\!\!\sqrt{3}k_zk_x$, and $d_5(\bm k)\!\!=\!\!\sqrt{3}k_xk_y$. The Dirac matrices, $\!\!\gamma_i$, are explicitly given as $\gamma_1\!=\!\sigma^x\otimes\mathbb{I},\gamma_2\!=\!\sigma^z\otimes\sigma^z,\gamma_3\!=\!\sigma^z\otimes\sigma^y,\gamma_4\!=\!\sigma^z\otimes\sigma^x$, and $\gamma_5\!=\!\sigma^y\otimes\mathbb{I}$ where $\sigma^{\alpha}$ are the Pauli matrices and $\mathbb{I}$ is the $2\!\times\!2$ identity matrix. It is worth to note that Eq.\eqref{eq:LH_c} is a complete representation of kinetics in Luttinger semimetal when both inversion symmetry and time-reversal symmetry are present. In Eq.\eqref{eq:LH_c}, $c_0$ quantifies the particle-hole asymmetry of the model, whereas, $c_i$ quantify the kinetic term proportional to each of the $d$-wave harmonics. 
When all $c_{i}$ are the same, the model in Eq. \eqref{eq:LH_c} becomes fully spherical symmetric retaining $SO(3)$ symmetry. In the case of cubic symmetry, whereas, we have $c_{1,2}\!\neq\!c_{3,4,5}$\cite{savary2017superconductivity}.

We now discuss the superconductivity emerging from this multiband Luttinger semimetal when the electron-electron interactions are present\cite{boettcher2018unconventional},
\bea
H_I\!=\!g_0(\psi^\dagger\psi)^2+\sum_a g_a(\psi^\dagger\gamma_a\psi)^2.
\label{eq:HI}
\eea
Using the Fierz identity, $H_I$ can be exactly rewritten in terms of the $s$-wave and $d$-wave pairing channels,
$H_I\!=\!H_s+\sum_{a}H_{d_{a}}$
with
\bea
H_s&\!=\!&g_s(\psi^{\dagger}\gamma_{45}\psi^{*})(\psi^{T}\gamma_{45}\psi), \nonumber \\
H_{d_{a}}&\!=\!&g_{d_a}(\psi^{\dagger}\gamma_a\gamma_{45}\psi^{*})(\psi^{T}\gamma_{45}\gamma_a\psi),
\label{eq:HI_s,d}
\eea
where $g_s\!=\!\frac{1}{4}(g_0+\sum_ag_a)$, $g_{d_{a}}\!=\!\frac{1}{4}(g_0+g_a-\sum_{b\neq a}g_b)$, and $\gamma_{45}\!=\!i\gamma_4 \gamma_5$. 
It is remarkable that the repulsive electron interaction with coefficients $g_a > 0$ can naturally induce the $d$-wave pairing instabilities.\cite{boettcher2018unconventional} For simplicity, we set coefficients $g_a\!=\!g_1$ and hence, $g_s\!=\!\frac{1}{4}(g_0+5g_1)$ and $g_{d_a}\!=\!\frac{1}{4}(g_0-3g_1)$.
In this work, we assume that the $d$-wave pairing channel is attractive such that  $g_{d_a}\!=\!-g$ and neglect $H_s$.
Within the standard mean-field decomposition, $H_I$ can be rewritten as follows up to the constant terms: 
\bea
H_I\!=\!-g\sum_a\big\{(\psi^{\dagger}\gamma_a\gamma_{45}\psi^{*})\Delta_a+(\psi^{T}\gamma_{45}\gamma_a\psi)\Delta_a^*\big\} , 
\label{eq:HI_m}
\eea
where the superconducting order parameters are explicitly given as
\bea
\Delta_{a}&\!=\!&\langle \psi^T_{-\bm k}\gamma_{45}\gamma_a\psi_{\bm k}\rangle.
\label{eq:sceq}
\eea
The order parameter $\Delta_a$ with $a \!\in \!( 1,2,\cdots 5)$ represents the $d$-wave quintet pairings ($j$ = 2). 
In particular, ${\Delta}_{e_g}\!\!\equiv\!\!(\Delta_1,\Delta_2)$ represents the two $d$-wave pairings, ($d_{x^2-y^2},d_{3z^2-r^2}$) with $e_g$ symmetry, and ${\Delta}_{t_{2g}}\!\!\equiv\!\!(\Delta_3,\Delta_4,\Delta_5)$ represents the three $d$-wave pairings, ($d_{yz},d_{zx},d_{xy}$) with $t_{2_g}$ symmetry.
Throughout our study, we consider the specific parameter set, which is relevant to PrBi, in Eq.\eqref{eq:LH_c} and analyze the properties of superconducting states; $c_0\!\!=\!\!-6(a/\pi)^2$eV, $c_{e_g}\!\!\equiv\!\!c_1\!\!=\!\!c_2\!\!=\!\!-2(a/\pi)^2$eV, and $c_{t_{2g}}\!\!\equiv\!\!c_3\!\!=\!\!c_4\!\!=\!\!c_5\!\!=\!\!-1(a/\pi)^2$eV with the lattice constant $a$ and the chemical potential $\mu\!\!=\!\!-0.6$eV, for the cubic symmetric case. Here and below, we consider the case where there are two distinct doubly degenerate Fermi surfaces for $\mu<0$ (normal band structure). Although we focus on the specific parameter set, we emphasize that similar argument holds for different cases and the emergence of complex superconducting states due to intertwined multipolar order is a generic feature.

\begin{figure*}[t!]
\begin{widetext}
\begin{center}
\includegraphics[width=1.5\columnwidth]{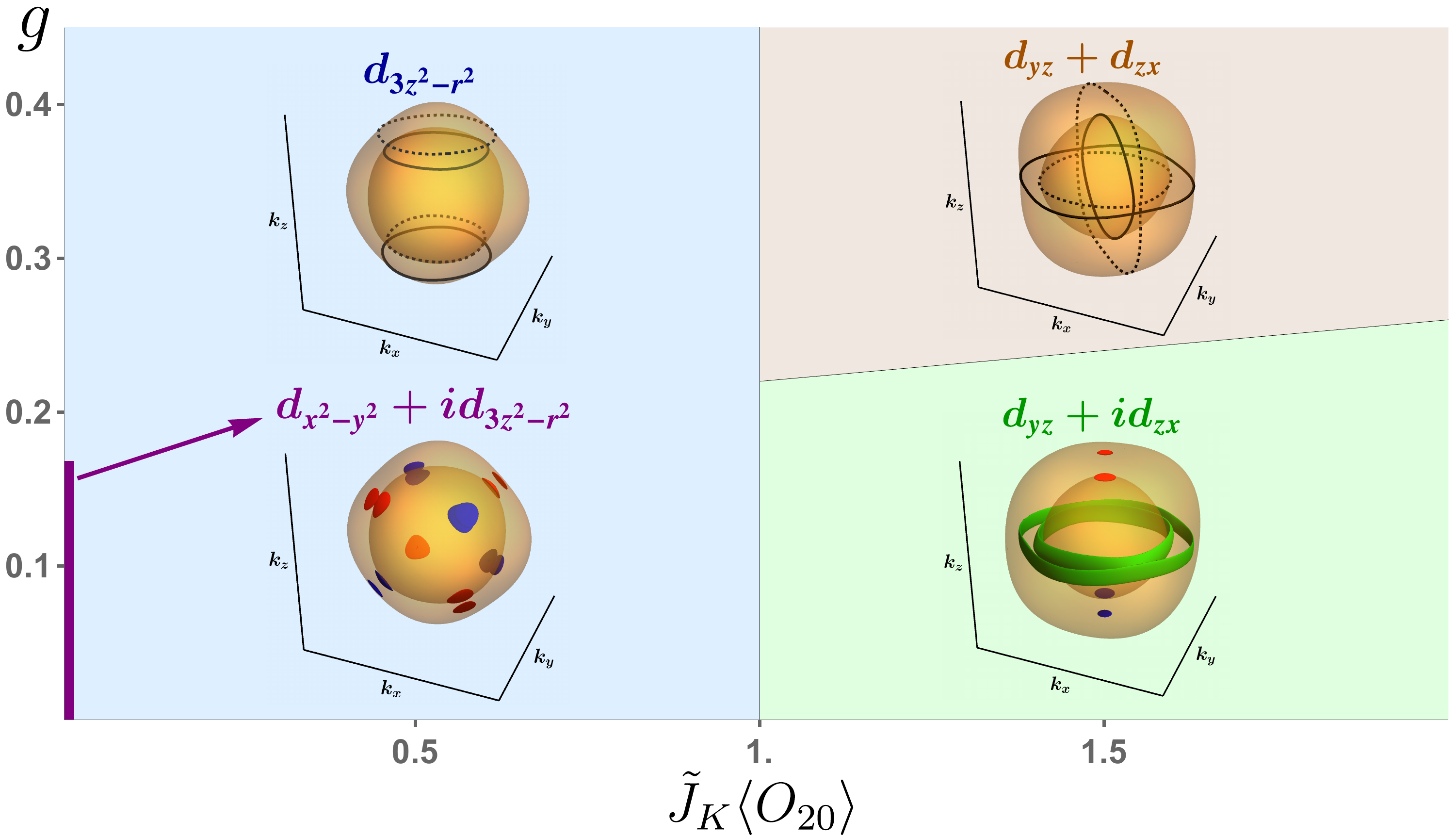}
\caption{(color online) Multipolar superconducting phases as functions of $\tilde{J}_{K} \langle O_{20} \rangle $ [$(a/\pi)^2$eV] and the interaction strength $g$ [eV] with $c_0\!\!=\!\!-6(a/\pi)^2$eV, $c_{e_g}\!\!=\!\!c_1\!\!=\!\!c_2\!\!=\!\!-2(a/\pi)^2$eV, $c_{\eta}\!\!=\!\!c_3\!\!=\!\!c_4\!\!=\!\!(-1-\tilde{J}_K\langle O_{20} \rangle)(a/\pi)^2$eV, $c_5\!\!=\!\!-1(a/\pi)^2$eV, and $\mu\!\!=\!\!-0.6$eV based on Eq.\eqref{eq:LH_c} and Eq.\eqref{eq:kondo-Jeff}. We find four distinct superconducting states: time-reversal symmetry breaking state with $d_{x^2-y^2}+id_{3z^2-r^2}$ pairing (purple line), $d_{yz}\!+\!id_{zx}$ pairing (blue), time-reversal symmetric state with $d_{3z^2-r^2}$ pairing (green), and $d_{yz}\!+\!d_{zx}$ pairing (brown). Four insets show the gap structure of these states. Note that the $\langle O_{20} \rangle\!=\!0$ vertical thick line corresponds to the case with the cubic symmetry in the absence of quadrupolar order. Here, each semi transparent yellow surface represents the normal state Fermi surface. For multipolar superconductor with time-reversal symmetry breaking, the Bogoliubov Fermi surfaces exist. Red, green and blue colored Bogoliubov Fermi surfaces indicate distinct Chern numbers $2$, $0$, and $-2$ respectively. For multipolar superconductor with time-reversal symmetry, each solid (dashed) ring indicates the nodal ring, which is protected by the non-trivial winding number $2$ $(-2)$. See the main text for more details.}
\label{fig:pd}
\end{center}
\end{widetext}
\end{figure*}

We first briefly discuss the superconducting phases in the absence of coexisting quadrupolar order, with the cubic symmetric Luttinger model where the coefficients in Eq.\eqref{eq:LH_c} are of the form, $|c_{e_g}|\!\!>\!\!|c_{t_{2g}}|$. In general, the free energy for the ${\Delta}_{e_g}$ pairing state is given as $F_{e_g}\!=\!r_{e_g}|\Delta_{e_g}|^2+q_1|\Delta_{e_g}|^4+q_2(\Delta_1\Delta_2^*-\Delta_2\Delta_1^*)^2$ while the free energy for the ${\Delta}_{t_{2g}}$ pairing state is given as  $F_{t_{2g}}\!=\!r_{t_{2g}}|\Delta_{t_{2g}}|^2+q'_1|\Delta_{t_{2g}}|^4+q'_2|\Delta_{t_{2g}}\!\cdot\Delta_{t_{2g}}|^2+q'_3(|\Delta_{3}|^2|\Delta_{4}|^2+|\Delta_{4}|^2|\Delta_{5}|^2+|\Delta_{5}|^2|\Delta_{3}|^2)$.\cite{sigrist1991phenomenological} Within one-loop calculation, the instability towards the ${\Delta}_{e_{g}}$ pairing is shown to be stronger than the ${\Delta}_{t_{2g}}$ pairing with $|c_{e_g}|\!\!>\!\!|c_{t_{2g}}|$, i.e., $r_{e_{g}}\!\!<\!\!r_{t_{2g}}$ (See Section I of Supplementary Information for details).
For the ${\Delta}_{e_{g}}$ pairing, there are three possible superconducting states with the order parameters $\Delta_{e_g}=(1,0), (0,1),$ and $(1,i)$. By comparing the mean-field energy at zero temperature, for weak coupling limit, i.e., small $g$ limit, we find that the time-reversal symmetry breaking superconducting phase is chosen, which is described by $d_{x^2-y^2} +i d_{3z^2-r^2}$ pairing or the order parameter $\Delta_{e_g} =(1,i)$. In this phase, the Bogoliubov quasiparticles form sixteen distinct pockets as shown in the bottom left inset of Fig.\ref{fig:pd}. Furthermore, we find that each pocket colored red (blue) is characterized by non-trivial Chern number $2$ $(-2)$, classified by $2\mathbb{Z}$ Chern number corresponding to Class D in the Altland-Zirnbauer classification with additional inversion symmetry\cite{bzduvsek2017robust}.

With increasing interaction strength $g$, we observe the superconducting phase transition occurs from the $d_{x^2-y^2} +i d_{3z^2-r^2}$ pairing state to the time-reversal symmetric $d_{3z^2-r^2}$ pairing state. This phase transition can be understood as the effect of band flattening near the quadratic band touching point. More precisely, the electron interaction starts to dominate over the kinetic energy at large $g$ and the system behaves similarly to the case with small $\mu$ due to the band flattening near ${\boldsymbol k}\!=\!0$ which favors the $d_{3z^2-r^2}$ pairing state\cite{boettcher2018unconventional}. In Fig.\ref{fig:pd}, the vertical thick line at $\tilde{J}_K\langle O_{20} \rangle=0$ corresponds to
$d_{x^2-y^2} +i d_{3z^2-r^2}$ pairing and there is the phase transition to $d_{3z^2-r^2}$ beyond that,
as the interaction strength $g$ increases. The BdG energy spectrum of this phase possesses gapless nodal rings as shown in the top left inset of Fig.\ref{fig:pd}. In this time-reversal symmetric superconductor, the solid (dashed) nodal line is protected by a non-trivial winding numbers $2$ $(-2)$ which belongs to the $2\mathbb{Z}$ classification of DIII Class\cite{bzduvsek2017robust}.

{\bf \em Multipolar Kondo coupling---} 
When multipolar degrees of freedom are present in the system, one should consider an effective Kondo coupling between the localized multipolar moments and itinerant electrons. In this section, we consider the microscopic model focusing on PrBi and derive the multipolar Kondo coupling between the $e_g$-type quadrupolar moments in Pr$^{3+}$ and the strongly spin-orbit coupled electrons of Bi $6p$ orbitals. The $e_g$-type quadrupolar degrees of freedom in the cubic symmetric model is represented in terms of the Stevens operators $O_{22} \!=\! \frac{\sqrt{3}}{2} (J_x^2 -J_y^2) $ and $O_{20}\!=\! \frac{1}{2} (3J_z^2 -J^2)$ with the $\mu$-th component of total angular momentum $J_\mu$\cite{stevens1952matrix,lea1962raising}. Regarding PrBi, we consider the interpenetrating face-centered cubic (FCC) lattice system, where the quadrupolar degrees of freedom $O_{22}$ and $O_{20}$ of the localized electrons reside in one FCC lattice and the itinerant electrons with $p$ orbitals reside in another FCC lattice (For details, see Fig.\ref{fig:kondo} in Supplementary Information). Then, one can write down the effective Kondo coupling between the quadrupolar order parameters $O_{22}$ and $O_{20}$ and the itinerant $p$ electrons as the following,
\begin{equation}
H_K \!=\!  J_K \! \sum_{\langle i,j \rangle} \sum_{a,\alpha} \Big( O_{22} \Gamma_{1, ij}^{a} c_{i a \alpha}^\dagger c_{j a \alpha} + O_{20} \Gamma_{2, ij}^{a} c_{i a \alpha}^\dagger c_{j a \alpha} \Big).
\label{eq:Kondo}
\end{equation}
Here, $c^\dagger_{i a \alpha}$ and $c_{i a \alpha}$ are the electron creation and annihilation operators at site $i$ with orbital $a \!\in\! (x,y,z)$ and spin $\alpha \!\in\! (\uparrow, \downarrow)$. $\Gamma_{1,ij}^{a }$ and $\Gamma_{2,ij}^{a }$ are site- and orbital-dependent form factors for the Kondo coupling with quadrupoles $O_{22}$ and $O_{20}$ respectively (See Section II of Supplementary Information for details). We note that the quadrupolar degrees of freedom which is time-reversal symmetric, can only couple to the spin independent electron hoppings with the form factors that transform exactly the same as $O_{22}$ and $O_{20}$. Now we take into account $j\!=\!3/2$ basis in the presence of the spin-orbit coupling of $p$ electrons and project Eq.\eqref{eq:Kondo} onto $j\!=\!3/2$ basis with the projection operator $P_{j=3/2}$\cite{stamokostas2018mixing}. Then one gets the following Kondo coupling, 
\bea
\tilde{H}_K ({\boldsymbol k}) &\!=\!& P_{j=3/2} H_{K} ({\boldsymbol k}) P_{j=3/2}   \nonumber \\
&\!=\!&  \tilde{J}_K \Big( (\sqrt{3}O_{20}\!+\!O_{22})d_3 ({\boldsymbol k})\gamma_3  \nonumber \\
&\!+\!&(\sqrt{3}O_{20}\!-\!O_{22})d_4 ({\boldsymbol k})\gamma_4 - 2O_{22}d_5 ({\boldsymbol k})\gamma_5  
\Big),
\label{eq:kondo-Jeff} 
\eea
in four component spinor basis $\psi$ and with $\tilde{J}_K \equiv J_{K}(\frac{a}{\pi})^2$. In Eq.\eqref{eq:kondo-Jeff}, one can clearly see that $O_{20}$-type ferro-quadrupolar ordering breaks the three-fold rotation symmetry, while $O_{22}$-type ferro-quadrupolar ordering breaks both the three-fold and the four-fold rotation symmetries. Recent experiment on PrBi compound has confirmed $O_{20}$-type ferro-quadrupolar order $\langle O_{20} \rangle \not= 0$, which has also been discussed within the Landau theory analysis on symmetry grounds\cite{he2019prbi,lee2018landau,freyer2018two}. Thus, we focus on the case when $O_{20}$-type ferro-quadrupolar order is present, $\langle O_{20} \rangle \neq 0$ and $\langle O_{22} \rangle \!=\!0$.

{\bf \em Ferro-quadrupolar order and superconductivity  ---} 
When $\langle O_{20} \rangle \!\neq\! 0$, the symmetry of the system is lowered to $D_{4h}$ from $O_h$ group\cite{ruan2016symmetry,shao2017strain}. One can easily see from both Eq.\eqref{eq:LH_c} and Eq.\eqref{eq:kondo-Jeff} that ferro-quadrupolar order gives rise to anisotropies in coefficients, $c_3\!=\!c_4\! \neq \! c_5$ and results in Fermi surface distortion. In this case, the coefficient $c_{\eta}\!\!\equiv\!\!c_3\!\!=\!\!c_4$ is renormalized in Eq.\eqref{eq:LH_c} and the spontaneous Fermi surface distortion occurs via the effective Kondo coupling shown in Eq.\eqref{eq:kondo-Jeff}.  
In particular, when the quadrupolar order induces the Fermi surface distortion, we find that the properties of the $d$-wave superconductivity is dramatically changed. In Fig.\ref{fig:pd}, we plot the phase diagram within mean-field approximation as functions of $\tilde{J}_{K} \langle O_{20} \rangle$ and the interaction strength $g$ at zero temperature. With the onset of $O_{20}$-type ferro-quadrupolar order, the instability towards the $\Delta_2$ pairing is shown to be stronger than the $\Delta_1$ pairing, which is consistent with the result of one-loop calculation (See Section I of Supplementary Information for details). Thus the system prefers the $d_{3z^2-r^2}$ pairing with $\Delta_{e_g} = (0,1)$ in both weak and strong coupling limits. With further increase of $\tilde{J}_{K} \langle O_{20} \rangle$, however, the instability towards the $\Delta_\eta\!\!\equiv\!\!(\Delta_3,\Delta_4)$ pairing becomes stronger than the $\Delta_2$ pairing. In general, the free energy for the $\Delta_{\eta}$ pairing state is represented as\cite{mineev1999introduction},  
\bea
\nonumber
F_{\eta}&\!=\!&r_{\eta}|\Delta_\eta|^2+q_1|\Delta_\eta|^4+q_2|\Delta_\eta\Delta_\eta|^2+q_3(|\Delta_3|^4+|\Delta_4|^4).
\label{eq:free}
\eea
Once the instability of the $\Delta_{\eta}$ pairing gets stronger than the $\Delta_2$ pairing, the phase transition to 
the $\Delta_{\eta}$ pairing occurs. For weak coupling limit, the system develops time-reversal symmetry breaking superconductivity with the $d_{yz}+i d_{zx}$ pairing and the order parameters $\Delta_\eta=(1,i)$. This result is distinct from the cubic case, where the $d_{x^2-y^2} +i d_{3z^2-r^2}$ pairing with ${\Delta}_{e_g}\!=\!(1,i)$ is chosen. As shown in the bottom right inset of Fig.\ref{fig:pd}, the Bogoliubov quasiparticles form four Fermi surfaces along $k_z$ axis with the Chern number $\pm2$ and two Fermi surfaces located at $k_z=0$ with the Chern number $0$. With increasing $g$, the phase transition occurs favoring distinct superconducting phase with the $d_{yz}+d_{zx}$ pairing described by the order parameter $\Delta_\eta=(1,1)$. In this case, the time-reversal symmetry is recovered and the Bogoliubov quasiparticles form four nodal rings with the winding numbers $\pm2$ as shown in the top right inset of Fig.\ref{fig:pd}.

{\bf \em Discussion ---} 
We have studied exotic multipolar superconductors and their topological properties, which arise from the intertwined multipolar order 
and electron correlations in the Luttinger semimetal.  
Considering the electron Coulomb interaction as the dominant driving force for superconductivity, we have shown that the $d$-wave pairing channel in the pseudospin $j$ = 2 manifold becomes attractive and there exists special selection of the $d$-wave superconducting order parameters. When the quadrupolar order of localized moments coexists, we have demonstrated how it can change the superconducting phases of the Luttinger semimetals. In particular, we consider the $e_g$-type quadrupolar order $O_{20}$ and $O_{22}$ present in the cubic symmetric systems.
We derived the effective Kondo coupling between the quadrupolar local moments and conduction electrons via
the microscopic model with spin-orbit-coupled $p$ electrons and projecting it onto the pseudospin $j$ = 3/2 Luttinger Hamiltonian. 
It turns out that the onset of ferro-quadrupolar order largely affects the Fermi surface distortion, and thereby 
causes dramatic changes in preferred superconducting order parameters. 
We emphasize that such phenomena are quite unique in the interacting Luttinger semimetals with relatively small carrier densities,
where the effective Kondo coupling with the quadrupolar degrees of freedom can sensitively control the nature of 
the superconducting order parameters and the associated topological properties.

Recent experiments on the semimetallic compound PrBi have confirmed the existence of $O_{20}$-type ferro-quadrupolar order below the transition temperature $T_Q$ = 0.08K\cite{he2019prbi}. In this material, the localized moments of Pr$^{3+}$ ions form a $\Gamma_3$ non-Kramers doublet via strong spin-orbit coupling, which only allows higher multipolar moments, but no dipole moment.
Whereas, the itinerant electrons of Bi $6p$ orbitals form a strongly correlated Luttinger semimetal with small carrier density\cite{vashist2019fermi, he2019prbi}. Since the system contains tiny carrier density, one may expect to control electron correlation via doping and external pressure, resulting in superconductivity driven by the interplay between the quadrupolar Kondo effect and the electron interaction. In such cases, as shown in Fig.\ref{fig:pd}, the multipolar superconductivity with distinct $d$-wave pairing order parameters are stabilized and depending on the presence and absence of ferro-quadrupolar order, the character of superconductivity and topological nature of the Bogoliubov quasiparticles may be sensitively changed. This can be verified by probing surface modes using scanning tunneling microscope. Moreover, for multipolar superconductors with time-reversal symmetry breaking pairing channels, the location of Bogoliubov Fermi surfaces with non-trivial Chern numbers can be sensitively changed, depending on distinct mixtures of the $d$-wave pairings, 
i.e. $d_{x^2-y^2} + i d_{3z^2-r^2}$ or $d_{yz}+i d_{zx}$ as shown in the bottom insets of Fig.\ref{fig:pd}. 
Thus, one expects strong angle dependence of the Hall effect signal, which would distinguish 
different superconducting phases.

With growing interest on multipolar order, often termed as ``hidden order", it is now known that there exist many systems, where both multipolar order and superconductivity may coexist. For instance, beyond the quadrupolar Kondo semimetal PrBi, the materials like rare-earth half-heusler compounds, Pr based cage compounds Pr(Ti,V,Ir)$_2$(Al,Zn)$_{20}$ and lacunar spinel compounds Ga(Ta,Nb)$_4$(S,Se)$_8$ contain 
spin-orbit entangled pseudospin degrees of freedom and sometimes exhibit (anti-) ferro-quadrupolar order in addition to superconductivity. In such cases, the multipolar Kondo coupling and strongly interacting multi-orbital electrons play an important role to determine the characteristics of superconductivity. Our results can be used to understand how these two phenomena can be intertwined with each other and how the topological properties of multipolar superconductors could be controlled via the multipolar order.
Our work provides an important platform for the discovery of magnetic topological superconductors that can be 
controlled by electron correlation or multipolar magnetism.

\acknowledgments
Y.B.K. is supported by the NSERC of Canada, Canadian Institute for Advanced Research, and Center for Quantum Materials at the University of Toronto. G.Y.C. is supported by BK21 plus program, POSTECH. A.M. is supported by BK21 plus. G.B.S., M.J.P., and S.B.L. are supported by the KAIST startup, BK21 and National Research Foundation Grant (NRF-2017R1A2B4008097). 

\bibliography{super_fermi_distortion_bib}

\renewcommand{\thefigure}{S\arabic{figure}}
\setcounter{figure}{0}
\renewcommand{\theequation}{S\arabic{equation}}
\setcounter{equation}{0}

\begin{widetext}
\section{Supplementary Information for ``Multipolar superconductivity in Luttinger semimetals''}

\section{Ginzburg-Landau Free Energy and one-loop expansion}

In this section, we compute the coefficient of the quadratic term, $r_a$, in the Ginzburg-Landau free energy $F(\Delta_a)$ to compare the strength of instabilities towards $\Delta_a$ pairing. We first introduce the free electron propagator
\bea
G(K)=(ik_0 + c_0k^2+\sum_{i}c_id_i(\boldsymbol{k})\gamma_i-\mu)^{-1}=\frac{-ik_0-c_0k^2 + \sum_{i}c_id_i(\boldsymbol{k})\gamma_i+\mu}{\sum_{i}(c_id_i(\boldsymbol{k}))^2-(c_0k^2+ik_0-\mu)^2}.
\eea
Here $K\equiv(k_0,\boldsymbol{k})$ and $k_0=2\pi(n+1/2)T$ denotes the Matsubara frequency. Then, the free energy is written as,
\bea
F(\vec\Delta)=\frac{1}{g} |\vec\Delta|^2 + T\sum_{m,n}\int_{\boldsymbol{k}}^{\Lambda} \frac{1}{m} \text{tr}(-G(K)\hat\Delta G(-K)^T\hat\Delta^\dagger)^m,
\eea
where $\hat\Delta=\sum_{a} \gamma_a\gamma_{45}\Delta_a$. Let $F_{2}(\Delta_a)$ be the contribution to the free energy that contains 2nd power of $\Delta_a$. We have
\bea
F_2(\Delta_{a})&=&\frac{1}{g} |\Delta_{a}|^2 -\frac{1}{2}L_{a}\Delta^*_{a}\Delta_{a}
\label{eq:f}
\eea
with
\bea
\nonumber
L_{a}&=&T\sum_{k_0}\int_{\boldsymbol{k}}^{\Lambda} \text{tr}(G(K)\gamma_aG(-K)\gamma_a)
\eea
which is represented as a Feynman diagram shown in Fig.\ref{fig:fd}. In this expression, we use the relation $\gamma_{45}G(K)^{\text{T}}\gamma_{45}=G(K)$.

\begin{figure}[h]
\centering\includegraphics[width=0.2\textwidth]{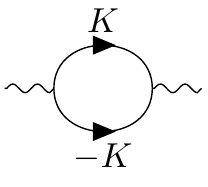}
\caption{Diagrammatic representation of $L_{a}$. Each solid arrow refer free electron propagator, $G(K)$, with $K=(k_0,\boldsymbol{k})$ while each wiggly line indicate insertion of $\Delta_a$ with vertex $\gamma_a$.}
\label{fig:fd}
\end{figure}

Meanwhile, we can parametrize the terms in free energy accordingly.
\bea
\nonumber
F_2(\Delta_a)&=&r_{a} |\Delta_{a}|^2
\label{eq:free_s}
\eea
Choosing the specific configurations,
\bea
\Delta_{e_g}^1=(1,0),~
\Delta_{e_g}^2=(0,1),~
\Delta_{t_{2g}}^1=(1,0,0),~
\Delta_{t_{2g}}^2=(0,1,0),~
\Delta_{t_{2g}}^3=(0,0,1)
\label{eq:ansatz}
\eea
we apply Eq.\ref{eq:f} and
\bea
F_2(\Delta_{e_g}^1)=r_1,~
F_2(\Delta_{e_g}^2)=r_2,~
F_2(\Delta_{t_{2g}}^1)=r_3,~
F_2(\Delta_{t_{2g}}^2)=r_4,~
F_2(\Delta_{t_{2g}}^3)=r_5~
\eea
to get the coefficients $r_a$. Then one can write the coefficients as below with $\hat{k}_0=k_0/T=2\pi(n+1/2), \hat{c}_i=c_i/T$ and $\hat{\mu}=\mu/T$.
\bea
r_a=\frac{1}{g}+T^{1/2}\int_{\boldsymbol{k}}\sum_n\frac{2\big(\sum_i(\hat{c}_id_i(\boldsymbol{k}))^2-2(\hat{c}_ad_a(\boldsymbol{k}))^2-(\hat{c}_0k^2-\hat{\mu})^2-\hat{k}_0^2\big)}{\big[\sum_{i}(\hat{c}_id_i(\boldsymbol{k}))^2-(\hat{c}_0k^2-\hat{\mu}+i\hat{k}_0)^2\big]\big[\sum_{i}(\hat{c}_id_i(\boldsymbol{k}))^2-(\hat{c}_0k^2-\hat{\mu}-i\hat{k}_0)^2\big]}
\eea
Remarkably, $r_a\!-\!r_b$ can be simply expressed as the following,
\bea
r_{a}-r_{b}=T^{1/2}\int_{\boldsymbol{k}}\sum_n\frac{-4(\hat{c}_ad_a(\boldsymbol{k}))^2+4(\hat{c}_bd_b(\boldsymbol{k}))^2}{\big[\sum_{i}(\hat{c}_id_i(\boldsymbol{k}))^2-(\hat{c}_0k^2-\hat{\mu}+i\hat{k}_0)^2\big]\big[\sum_{i}(\hat{c}_id_i(\boldsymbol{k}))^2-(\hat{c}_0k^2-\hat{\mu}-i\hat{k}_0)^2\big]}.
\eea

In Fig.\ref{fig:mass}, we plot the numerical evolution of coefficients, $\tilde{r}_a \equiv r_a/T^{1/2}-1/g$, as a function of $\hat{O}_{20}\!\equiv\!\hat{c}_5-\hat{c}_\eta$ with $\hat{c}_0\!=\!-2000(a/\pi)^2$, $\hat{c}_{e_g} \!\equiv\! \hat{c}_1\!=\!\hat{c}_2\!=\!-2000/3(a/\pi)^2$, $\hat{c}_{\eta}\!\equiv\!\hat{c}_3\!=\!\hat{c}_4\!=\!(-1000/3-\hat{O}_{20})(a/\pi)^2$, $\hat{c}_5\!=\!-1000/3(a/\pi)^2$, and $\hat{\mu}\!=\!-200$ as appropriate for PrBi. First, it clearly shows that the instablity towards ${\Delta}_{e_{g}}$ pairing is stronger than ${\Delta}_{t_{2g}}$ pairing, i.e., $\tilde{r}_{e_{g}}\!\equiv\!\tilde{r}_{1}\!=\!\tilde{r}_{2}\!<\!\tilde{r}_{t_{2g}}\!\equiv\!\tilde{r}_{3}=\tilde{r}_{4}=\tilde{r}_{5}$, with $\hat{O}_{20}=0$ for cubic symmetry as stated in the main text. Moreover, Fig.\ref{fig:mass} also shows that the instability towards $\Delta_2$ pairing becomes stronger than $\Delta_1$ pairing, i.e., $\tilde{r}_{2}\!<\!\tilde{r}_{1}$, as soon as $O_{20}$-type ferro-quadrupolar order becomes finite, $\hat{O}_{20} \not= 0$. Finally, it tells us that the instability towards $\Delta_\eta\!\!\equiv\!\!(\Delta_3,\Delta_4)$ pairing becomes stronger than $\Delta_2$ pairing, $\tilde{r}_{\eta}\!\equiv\!\tilde{r}_3\!=\!\tilde{r}_4<\!\tilde{r}_{2}$, for $\hat{O}_{20}>\hat{c}_5-\hat{c}_{e_g}=1000/3(a/\pi)^2$.

\begin{figure}[h]
\centering\includegraphics[width=0.8\textwidth]{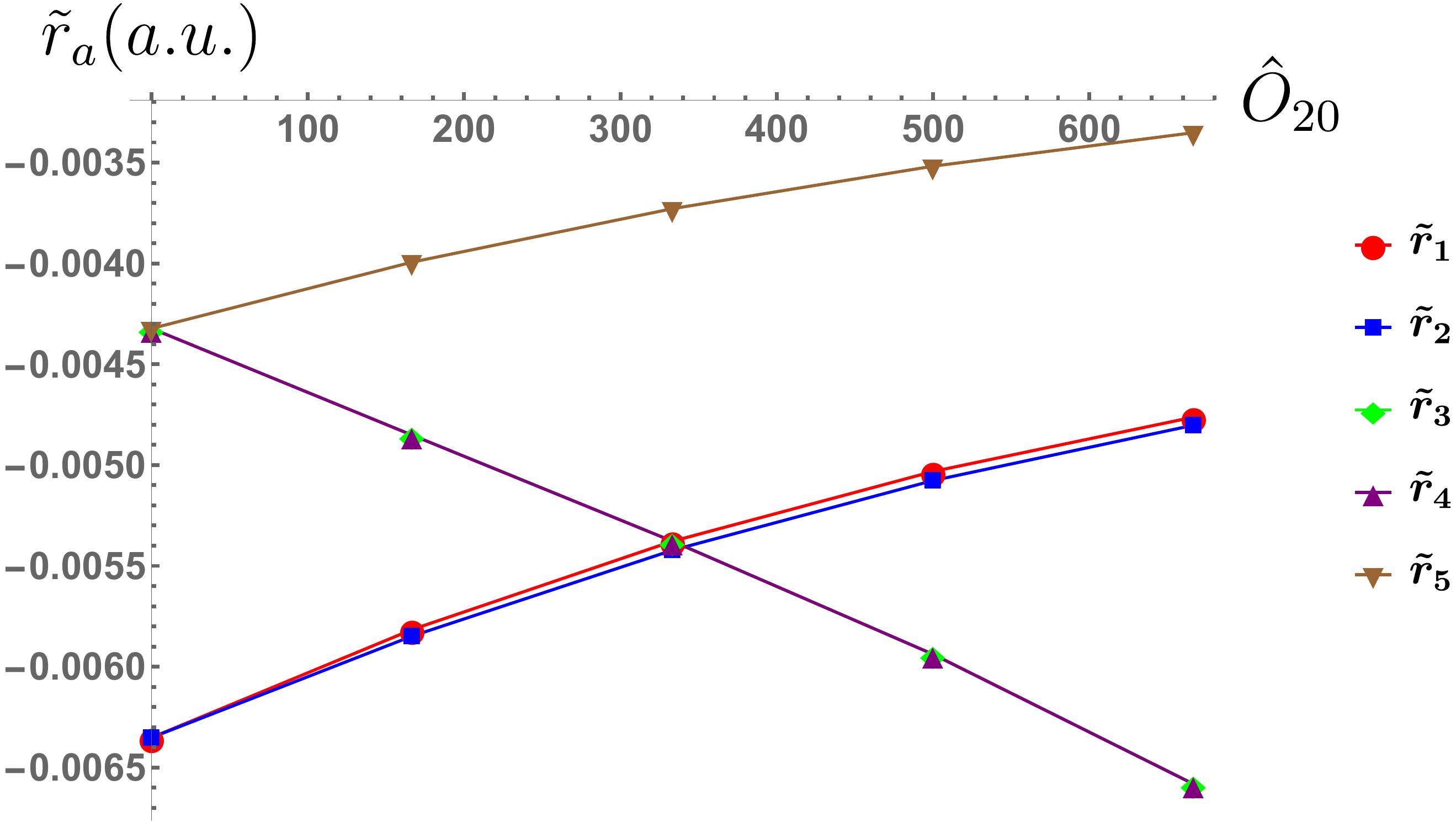}
\caption{(color online) Plot of coefficients, $\tilde{r}_a \equiv r_a/T^{1/2}-1/g$, as a function of $\hat{O}_{20}$ [$(a/\pi)^2$] with $\hat{c}_0\!=\!-2000(a/\pi)^2$, $\hat{c}_{e_g}\!\equiv\!\hat{c}_1\!=\!\hat{c}_2\!=\!-2000/3(a/\pi)^2$, $\hat{c}_{\eta}\!\equiv\!\hat{c}_3\!=\!\hat{c}_4\!=\!(-1000/3-\hat{O}_{20})(a/\pi)^2$, $\hat{c}_5\!=\!-1000/3(a/\pi)^2$, and $\hat{\mu}\!=\!-200$. These numbers are relevant to PrBi.}
\label{fig:mass}
\end{figure}

\section{Kondo coupling and Fermi surface distortion}

In this section, we derive the effective Kondo coupling between the quadrupolar order parameters and the itinerant $j=3/2$ electrons for the interpenetrating FCC lattice system. We start by introducing the Kondo model where the quadrupolar order parameters $O_{22}$ and $O_{20}$ and the itinerant $t_{2g}$ electrons couple as the following,
\begin{equation}
H_K \!=\!  J_K \! \sum_{\langle i,j \rangle} \sum_{a,\alpha} \Big( O_{22} \Gamma_{1, ij}^{a} c_{i a \alpha}^\dagger c_{j a \alpha} + O_{20} \Gamma_{2, ij}^{a} c_{i a \alpha}^\dagger c_{j a \alpha} \Big).
\label{eq:Kondo_s}
\end{equation}
Here, $c^\dagger_{i a \alpha}$ and $c_{i a \alpha}$ are the electron creation and annihilation operators at site $i$ with orbital $a \!\in\! (x,y,z)$ and spin $\alpha \!\in\! (\uparrow, \downarrow)$. We consider the case where the quadrupolar degrees of freedom from the localized electron reside in one FCC lattice and the itinerant electrons with $p$ orbitals reside in another FCC lattice as in Fig.\ref{fig:kondo}.
\begin{figure}[h]
\includegraphics[width=0.3\columnwidth]{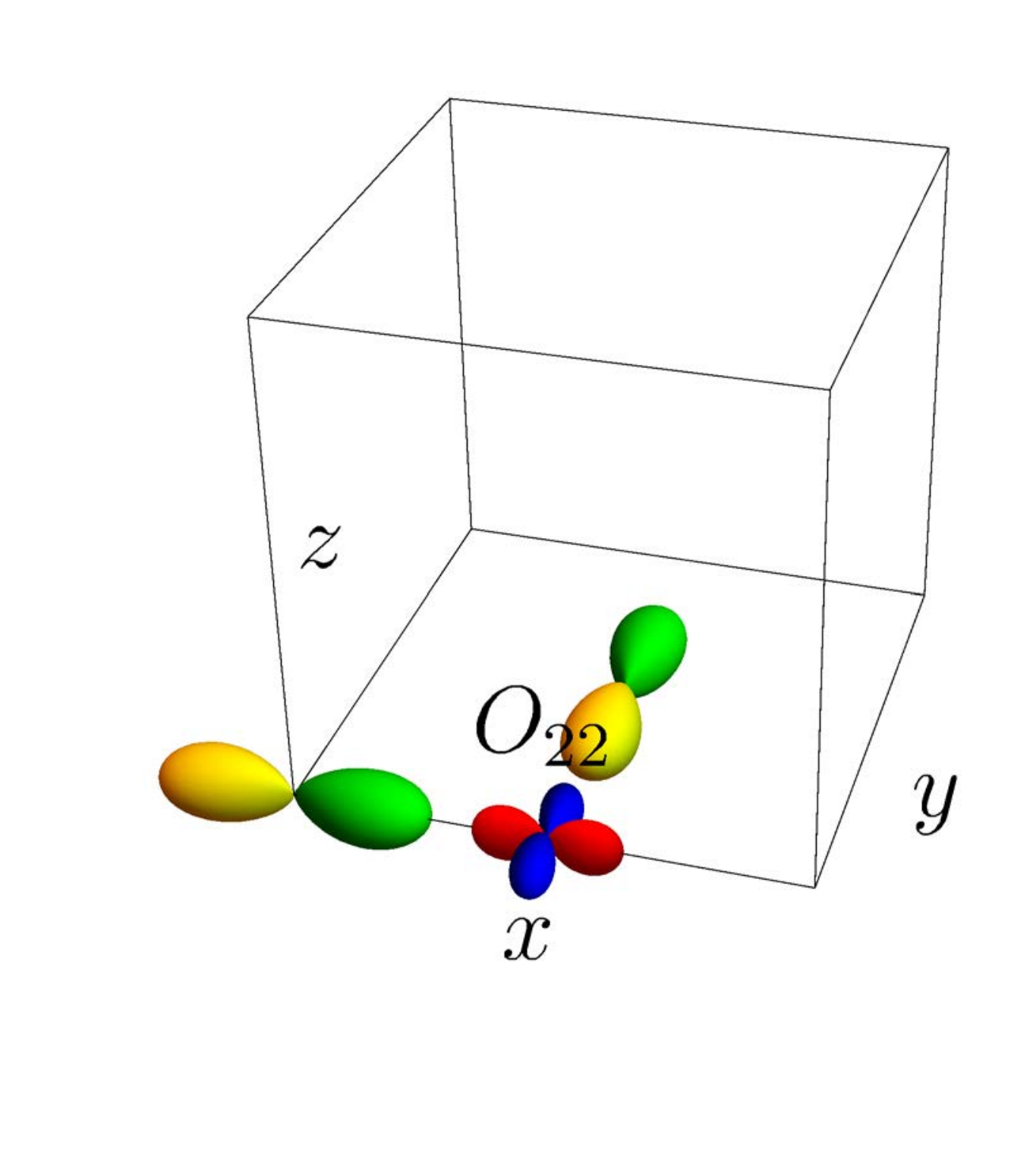}
\caption{(color online) Position of the quadrupolar degrees of freedom and the itinerant electrons with $p$ orbitals in the interpenetrating FCC lattice system. One of two quadrupole moments, $O_{22}$, is colored red and blue. $p_x$ orbital is colored green and yellow.}
\label{fig:kondo}
\end{figure}
Then one of the Kondo coupling terms, which couples itinerant electrons in $p_x$ orbital with other itinerant electrons in $p_y$ orbital residing on nearest neighbor sites, can be written as below.
\bea
H_{ixy}\!=\!J_K\sum_{\alpha}O_{22}(c_{i_x y \alpha}^\dagger c_{i x \alpha}-c_{i_y x \alpha}^\dagger c_{i y \alpha}+ c_{i_{-x} y \alpha}^\dagger c_{i x \alpha}-c_{i_{-y} x \alpha}^\dagger c_{i y \alpha})
\label{eq:Hxy}
\eea
where the site index $i_a$ represents the nearest neighbor of site $i$ in $a$ direction. Here, the $-$ sign for the second term comes from the $O_{22}$, which transforms as $O_{22} \to -O_{22}$ under $C_{4z}$
($\pi/2$ rotation about $z$ axis). Using $C_{31}$ rotation ($2\pi/3$ rotation along $(111)$ direction), we can write symmetry related terms as below,
\bea
\nonumber
H_{iyz}\!=\!J_K\sum_{\alpha}(-\frac{1}{2}O_{22}-\frac{\sqrt{3}}{2}O_{20})(c_{i_y z \alpha}^\dagger c_{i y \alpha}-c_{i_z y \alpha}^\dagger c_{i z \alpha}+ c_{i_{-y} z \alpha}^\dagger c_{i y \alpha}-c_{i_{-z} y \alpha}^\dagger c_{i z \alpha}),\\
H_{izx}\!=\!J_K\sum_{\alpha}(-\frac{1}{2}O_{22}+\frac{\sqrt{3}}{2}O_{20})(c_{i_z x \alpha}^\dagger c_{i z \alpha}-c_{i_x z \alpha}^\dagger c_{i x \alpha}+ c_{i_{-z} x \alpha}^\dagger c_{i z \alpha}-c_{i_{-x} z \alpha}^\dagger c_{i x \alpha}).
\eea
After Fourier transforming the Hamiltonian, $H_{K}=\sum_{i}(H_{ixy}+H_{iyz}+H_{izx})$, and expanding around $\bs{k}=0$, the Kondo Hamiltonian is written as
\bea
\nonumber
H_{K} ({\boldsymbol k})&=&J_K\bigg(\frac{a}{\pi}\bigg)^2\sum_{\bs{k}}\sum_{\alpha}\big((O_{22}(-k_x^2-k_y^2-k_xk_y)(c_{\bs{k} x \alpha}^\dagger c_{\bs{k} y \alpha} + c_{\bs{k} y \alpha}^\dagger c_{\bs{k} x \alpha})\\
\nonumber
&+&(-\frac{1}{2}O_{22}-\frac{\sqrt{3}}{2}O_{20})(-k_y^2-k_z^2-k_yk_z)(c_{\bs{k} y \alpha}^\dagger c_{\bs{k} z \alpha} + c_{\bs{k} z \alpha}^\dagger c_{\bs{k} y \alpha})\\
&+&(-\frac{1}{2}O_{22}+\frac{\sqrt{3}}{2}O_{20})(-k_z^2-k_x^2-k_zk_x)(c_{\bs{k} z \alpha}^\dagger c_{\bs{k} x \alpha} + c_{\bs{k} x \alpha}^\dagger c_{\bs{k} z \alpha})\big).
\eea 
By projecting onto $j=3/2$ basis with the projection operator $P_{j=3/2}$, one gets the following Kondo coupling,
\begin{eqnarray}
\tilde{H}_K ({\boldsymbol k}) &\!=\!& P_{j=3/2} H_{K} ({\boldsymbol k}) P_{j=3/2}   \nonumber \\
&\!=\!&  \tilde{J}_K \Big( (\sqrt{3}O_{20}\!+\!O_{22})d_3 ({\boldsymbol k})\gamma_3\!+\!(\sqrt{3}O_{20}\!-\!O_{22})d_4 ({\boldsymbol k})\gamma_4 - 2O_{22}d_5 ({\boldsymbol k})\gamma_5  
\Big)
\end{eqnarray}
in four component spinor basis $\psi$.

\end{widetext}

\end{document}